\documentclass[10pt, doublecolumn]{IEEEtran}
\usepackage{graphicx}
\usepackage{caption}
\usepackage[ruled]{algorithm2e}
\ifCLASSOPTIONcompsoc
  \usepackage[caption=false,font=normalsize,labelfont=sf,textfont=sf]{subfig}
\else
  \usepackage[caption=false,font=footnotesize]{subfig}
\fi
\usepackage{indentfirst}
\usepackage{graphicx}
\usepackage{bm}
\usepackage{amsmath}
\allowdisplaybreaks[4]
\usepackage{amssymb}
\usepackage{times}
\usepackage{mathtools}
\usepackage{psfrag}
\usepackage{cite}
\usepackage{lastpage}
\usepackage{fancyhdr}
\usepackage{color}
 \usepackage{amsthm}
\usepackage{bigints}
\newtheorem{Lemma}{Lemma}
\theoremstyle{remark}
\newtheorem{Remark}{$\quad$Remark}

\begin{document}
\title{A Stochastic Geometric Analysis on Multi-cell Pinching-antenna Systems under Blockage Effect}
\author{Yanshi Sun, \IEEEmembership{Member, IEEE}, Zhiguo Ding, \IEEEmembership{Fellow, IEEE}, and George K. Karagiannidis, \IEEEmembership{Fellow, IEEE}
\thanks{Y. Sun is with the School of Computer Science and Information
Engineering, Hefei University of Technology, Hefei, 230009, China. (email: sys@hfut.edu.cn.
Z. Ding is with the Department of Electrical and Electronic Engineering,
the University of Manchester, Manchester, M13 9PL, U.K. (e-mail:
zhiguo.ding@manchester.ac.uk).
G. K. Karagiannidis is with Department of Electrical and Computer Engineering, 
Aristotle University of Thessaloniki, Greece.(geokarag@auth.gr).
}\vspace{-3em}}
\maketitle
\begin{abstract}
Recently, the study on pinching-antenna technique has attracted significant attention. However, most relevant
literature focuses on a single-cell scenario, where the effect from the interfering 
pinching-antennas on waveguides connected to spatially distributed base stations (BSs) was ignored. To fulfill this 
knowledge gap,  this letter aims to provide an analytical framework on performance evaluation
for multi-cell pinching-antenna systems where spatially distributed waveguides which are connected to different 
BSs are considered. In particular, tools from stochastic geometry is applied for system modeling. 
The expression for 
the outage probability is obtained. Simulation results are provided to verify the accuracy of the analysis 
and demonstrate the superior performance of pinching-antenna system compared to fixed-antenna systems. 
\end{abstract}

\section{Introduction}
Recently, the research on flexible-antenna techniques,  which aims to provide the base stations (BSs) with the capability 
to re-build wireless channels between transmitters and receivers for performance enhancement, has attracted tremendous 
attention. To this end, a novel flexible-antenna technique termed ``pinching-antenna''
has recently been proposed \cite{suzuki2022pinching,ding2025flexible,yang2025pinching,liu2025pinching}. 
The key idea of pinching-antenna technique is to pinch separate dielectric materials onto waveguides which are fed to 
radio frequency chains so that the electromagnetic signals can be radiated from the pinched positions. 
Since the positions of the pinching-antennas can be adjustable, more flexibility can be provided for wireless transmission
design. The unique feature of pinching-antenna technique compared to other 
flexible-antenna techniques, such as reconfigurable intelligent surface (RIS), fluid-antenna systems (FAS) and movable-antenna (MA) technique, is that the pinching-antenna can provide large-scale reconfiguration of channels while 
others focus on re-building or leveraging small-scale fading \cite{ding2025flexible}. 

The first prototype system of pinching-antenna technique was provided by NTT DOCOMO \cite{suzuki2022pinching}. 
The performance enhancement provided by pinching-antenna compared to existing fixed-antenna systems 
was first analyzed by \cite{ding2025flexible}. Practical implementation issues on pinching-antenna techniques 
and the application of pinching-antenna techniques to advanced wireless communication techniques has been studied, showing pinching-antenna
as an emerging and potential research topic \cite{Kaidi24pinchingact,Xu24pinchinga,Chenguangji2025pinching,qin2025joint}. 

However, almost all the existing work on pinching-antenna systems focuses on a single-cell scenario by merely taking 
into consideration  one or multiple waveguides connected to a single BS, which 
ignores the impact of the interferences from nearby cells and hence may cause mismatch to the performance in reality.
Therefore, a natural question arises: what is the performance of pinching-antenna systems 
by considering the interferences from pinching-antennas on waveguides connected to different BSs which are 
spatially distributed. To this end, this paper aims to provide an analytical framework on performance evaluation
for multi-cell pinching-antenna systems where spatially distributed waveguides which are connected to different 
BSs are considered. In particular, tools from stochastic geometry is applied to model the locations 
of the nodes in pinching-antenna systems, and the effect of blockage is considered. The expressions for the outage 
probability achieved by the pinching-antenna system is obtained, which shows significant performance improvement compared to 
fixed-antenna systems. 
\begin{figure}[!t]
  \centering
      \setlength{\abovecaptionskip}{0em}   % ����ͼƬ������ͼƬ����
      \setlength{\belowcaptionskip}{-2em}   % ����ͼƬ���������ľ���
  \includegraphics[width=2.8in]{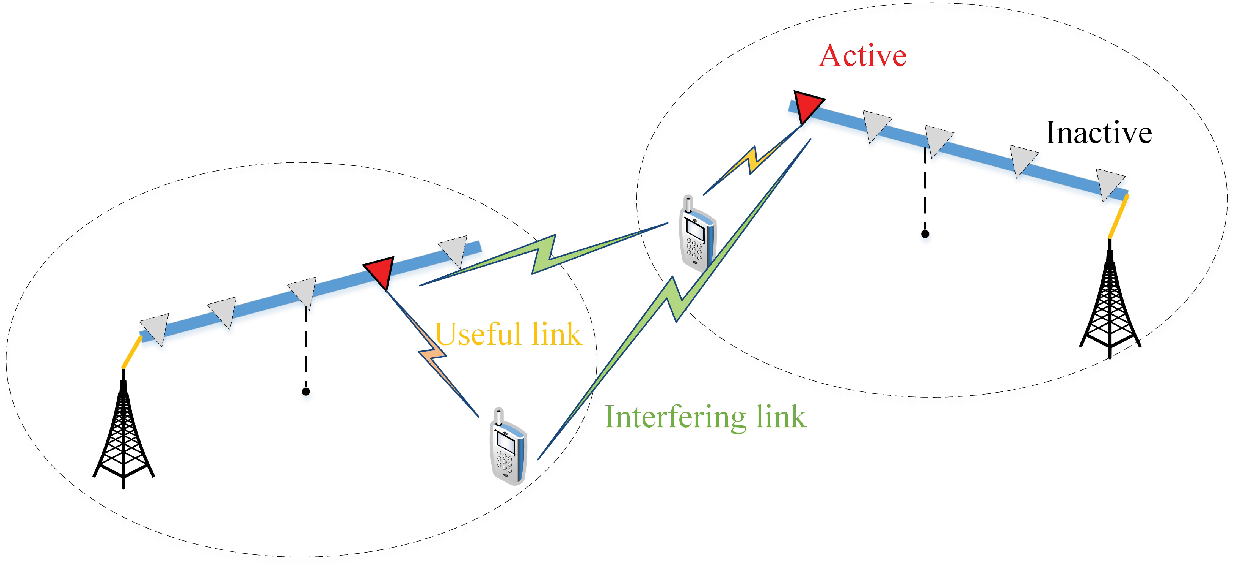}\\
  \caption{Illustration of the multi-cell pinching-antenna system.}\label{system_model}
\end{figure}

\section{System Model}
Consider a multi-cell pinching-antenna system as shown in Fig. \ref{system_model}, where the locations of the users are clustered which can be modeled
as a Poisson cluster point process (PCPP) denoted by $\Phi_c$ \cite{haenggi2012stochastic}. Specifically, the centers of the clusters 
(each with radius $\mathcal{R}$) form
a Poisson point process (PPP) with intensity $\lambda$ denoted by $\Phi_p=\{\mathbf{z}_i, i\geq 0\}$ with $\mathbf{z_i}=(x^c_i, y^c_i)$ being the coordinate of 
the center of the $i$-th cluster. In each cluster, there are $K_u$ independently and uniformly
distributed users in each cluster.  Different from existing fixed-antenna systems where the locations 
antennas of the BSs are fixed at the center of each cluster with height $H$, pinching-antenna assisted BS is considered 
in this paper where the locations of the antennas can be changed dynamically to provide more flexible 
re-configurations at the BS side. In particular, each BS is fed with a waveguide with length $L_w$ ($L_w/2<\mathcal{R}$) which is centered
at the center of each cluster also with height $H$. Note that the orientation denoted by $\theta_i$ of the waveguide of the $i$-th BS is uniformly distributed in 
$[0,\pi)$. Moreover, each waveguide is pinched with a pinching antenna, where the electromagnetic signal can be radiated. 
Particularly, this paper considers a practical method for implementing pinching-antenna systems, where 
the pinching-antenna  can only be activated at the one of the $\bar{N}_{P}$ preset locations \cite{Kaidi24pinchingact}. For simplicity, 
$\bar{N}_{P}$ is assumed to be an odd number in this paper and the plane coordinate of the $n$-th preset location on the waveguide
of the $i$-th BS is given by 
\begin{align}
\mathbf{\psi}_{i,n}=
(x_{i,n}^{\text{PA}},y_{i,n}^{\text{PA}})=&\left(\frac{L}{\bar{N}_{P}-1}(n-\frac{\bar{N}_{P}+1}{2})
\cos(\theta_i)+x^c_i.\right.\\\notag
&\left.\frac{L}{\bar{N}_{P}-1}(n-\frac{\bar{N}_{P}+1}{2})\sin(\theta_i)+y^c_i\right),
\end{align}

Orthogonal multiple access (OMA) is assumed in this paper, so that in each channel resource block
there's only one user served by the pinching-antenna assisted BS in each cluster. In addition, it is assumed that 
a user is served by the pinching antenna which is activated at the nearest pre-set location to the user, 
and the index of the chosen position is denoted by $n_i^*$ for the $i$-th cluster. 

Without loss of generality, consider a typical cluster, termed as the $0$-th cluster,  
whose center is at the origin and the serving waveguide is parallel
to the x-axis, which is reasonable due to the isotropic property of the considered 
geometric model in this paper. Then the locations of the interfering 
pinching-antennas can be denoted by $\tilde{\Phi}=\{\mathbf{\psi}_{i,n_i^*}, i\geq 1\}$

Consider a typical user in the $0$-th cluster whose location is denoted by $\mathbf{z}_0^{U}$, the received signal at the served user in the typical cluster can be given by:
\begin{align}
  r_0=\sqrt{\eta P} h_0s_0+\sum_{i=1}^{\infty}\sqrt{\eta P} h_is_i+n_0,
\end{align} 
where $s_0$ is the signal intended for the served user of the typical cluster 
and  $s_i$ ($i\geq 1$) is the interfering signal transmitted by the $j$-th pinching-antenna BS, $P$ is the transmitting power,
$n_0$ is background additive Gaussian noise whose power is $\sigma^2$, 
$\eta=\frac{c^2}{16\pi^2f_c^2}$, $c$ denotes the speed of light, $f_c$ is the carrier frequency
and $h_j$ is channel gain from the $j$-th BS which is given by 
\begin{align}
  h_j={g_j}{d_j^{-\frac{\alpha_j}{2}}}e^{-j\frac{2\pi||\psi_{j,n_j^*}-\psi_{j,1||}}{\lambda_g}},
\end{align}
where $g_j$ is the small-scale fading which is modeled as a Nakagami fading, 
thus $|g_j|^2$ follows a normalized Gamma distribution with an integer scale parameter $M_j$ and a scale parameter
$\frac{1}{M_j}$, $d_j$ is the distance from the pinching antenna of the $j$-th BS, $\alpha_j$ is the 
large-scale path loss exponent, $e^{-j\frac{2\pi||\psi_{j,n_j^*}-\psi_{j,1||}}{\lambda_g}}$ is the in-waveguide phase shift. 
Note that this paper takes into consideration the effect of blockage for channel modeling. Specifically, 
$h_j$ can be classified into either a line-of-sight (LoS) link or a non-line-of-sight (NLoS) link, depending on 
whether the link from the pinching antenna to the user is blocked by surroundings. Based on the rectangle Boolean modeling
method, $h_j$ belongs to LoS with probability $P_{L}(d_j)=\exp{(-\beta d_j)}$, where $\beta$ is a parameter related
to the properties of the blockages, and NLoS with probability  $P_N=1-P_{L}(d_j)$. Furthermore, for a LoS link, 
the small-scale parameter $N_j$ and large-scale parameter $\alpha_j$ are given by $N_j=N_L$ and $\alpha_j=\alpha_L$, 
where $N_L$ and $\alpha_L$ are constant; while for 
a NloS link, $N_j=N_N$ and $\alpha_j=\alpha_N$, where  $N_N$ and $\alpha_N$ are constant.

Therefore, the signal-to-interference-plus-noise ratio (SINR) for decoding $s_0$ can be given as follows:
\begin{align}
 \text{SINR}=\frac{|h_0|^2}{I+\xi},
\end{align}
where $I=\sum_{i=1}^{\infty}|h_i|^2$ denotes the sum power of the interferences normalized by $\eta P$, $\xi=\frac{\sigma^2}{\eta P}$. 

\section{Performance Analysis}
In this section, the outage probability achieved by the typical user will be provided, which is defined as follows:
\begin{align}
 P_{out}=\mathbb{P}\left(\log(1+\text{SINR}<\bar{R})\right),
\end{align}
where $\bar{R}$ is the target data rate of the user. 

\subsection{Laplace transform of the interferences}
To derive the expression for the outage probability, it is necessary to first characterize 
the Laplace transform of the interference $I$, denoted by $\mathcal{L}_I(s)=\mathbb{E}\{e^{-sI}\}$, which is highlighted
as follows. 

\begin{Lemma}
  The Laplace transform of $I$ can be approximated as follows: 
  \begin{align}\label{Lap}
   &\quad\mathcal{L}_I(s)\\\notag 
   &\approx\exp\!\bigg(\!\!-\frac{\pi^3\lambda}{2K}\!\!\!\!\!\sum_{Q\in\{L,N\}}\sum_{k=1}^{K}
   \sqrt{1-\theta_k^2}P_Q\left(\sqrt{\tan^2{\phi_k}+H^2}\right)\\\notag
   &\quad\quad\frac{\sin \phi_k}{\cos^3 \phi_k}
   \bigg(1-{\left(1\!+\!\frac{s}{N_Q \left(\tan^2{\phi_k}+H^2\right)^{\frac{\alpha_Q}{2}} }\right)^{-N_Q}}\bigg)\bigg),
  \end{align} 
where $\theta_k=\cos{\frac{(2k-1)\pi}{2K}}$, $\phi_k=\frac{\pi}{4}(1+\theta_k)$, and $K$ is the parameter of the 
Gaussian-Chebyshev approximation.
\end{Lemma}

\begin{IEEEproof}
$\mathcal{L}_I(s)$ can be written as follows:
\begin{align}
\mathcal{L}_I(s)&=\mathbb{E}_{\tilde{\Phi}}\left\{\prod_{i=1}^{\infty}\mathbb{E}_{|h_i|^2}
\left\{e^{-\frac{s|g_i|^2}{d_i^{\alpha_i}}}\right\}\right\}\\\notag
&=\mathbb{E}_{\tilde{\Phi}}\left\{\prod_{i=1}^{\infty}\sum_{Q\in\{L,N\}}
P_Q(d_i)\left(1+\frac{s}{N_Q d_i^{\alpha_Q}}\right)^{-N_Q}
\right\},
\end{align}
which follows from the fact that the $|g_i|^2$s are independently distributed 
normalized Gamma distributions. 

According to Slivnyak's theorem \cite{haenggi2012stochastic}, conditioning on that there's a point, i.e. $\mathbf{z}_0\in\Phi_P$,
then the remaining points $\Phi_{P \backslash \mathbf{z}_0}$ is also a PPP with intensity $\lambda$.
Furthermore, by noting that $\tilde{\Phi}$ is obtained by independently displacing 
all the points of $\Phi_{P} \backslash \mathbf{z}_0$, it can be concluded that $\tilde{\Phi}$ also forms 
a PPP with intensity $\lambda$. Moreover, consider a new point process which is given by 
$\tilde{\Phi}_{\mathbf{z_0}}=\tilde{\Phi}+\mathbf{z_0}$,
it can be obtained that $\tilde{\Phi}_{\mathbf{z_0}}$ is also a PPP with intensity $\lambda$. 
Then, $\mathcal{L}_I(s)$ can be further evaluated as follows:
\begin{align}
&\quad\mathcal{L}_I(s)\\\notag
&=\!\!\mathbb{E}_{\tilde{\Phi}_{\mathbf{z}_0}}\left\{\!\!\prod_{\mathbf{x}\in\tilde{\Phi}_{\mathbf{z}_0}}^{\infty}\sum_{Q\in\{L,N\}}
\!\!\!\!P_Q(D(\mathbf{x}))\left(1\!+\!\frac{s}{N_Q (D(\mathbf{x}))^{\alpha_Q}}\right)^{-N_Q}\right\}\\\notag
&\overset{(a)}{=}\exp\bigg(\!\!-\!\lambda\int_{\mathbb{R}^2}\left(1-\!\!\sum_{Q\in\{L,N\}}
\!\!\frac{P_Q(D(\mathbf{x}))}{\left(1\!+\!\frac{s}{N_Q (D(\mathbf{x}))^{\alpha_Q}}\right)^{N_Q}}\right)\ dx\bigg)\\\notag
&\overset{(b)}{=}\exp\!\bigg(\!\!-\!\lambda\!\!\!\!\!\sum_{Q\in\{L,N\}}\int_{\mathbb{R}^2}\bigg(1-
\!\!\frac{1}{\left(1\!+\!\frac{s}{N_Q (D(\mathbf{x}))^{\alpha_Q}}\right)^{N_Q}}\bigg)\\\notag 
&\quad \quad \quad \quad P_Q(D(\mathbf{x}))\ dx \bigg),
\end{align}
where $D(\mathbf{x})=\sqrt{||x||^2+H^2}$, step (a) follows by applying the probability generating functional (PGFL) of the PPP, and step (b) follows from the
fact the $P_L(||x||)+P_N(||x||)=1$. 

By changing to polar coordinates,  $\mathcal{L}_I(s)$ can be written as:
\begin{align}
&\quad\mathcal{L}_I(s)\\\notag
&=\exp\!\bigg(\!\!-2\pi\lambda\!\!\!\!\!\sum_{Q\in\{L,N\}}\int_{0}^{\infty}\bigg(1-
\!\!\frac{1}{\left(1\!+\!\frac{s}{N_Q \left(r^2+H^2\right)^{\frac{\alpha_Q}{2}}}\right)^{N_Q}}\bigg)\\\notag 
&\quad \quad \quad \quad P_Q\left(\sqrt{r^2+H^2}\right)r\ dr \bigg).
\end{align}
Then, by setting $t=arctan(r)$,  $\mathcal{L}_I(s)$ can be written as: 
\begin{align}
&\quad\mathcal{L}_I(s)\\\notag
&=\exp\!\bigg(\!\!-2\pi\lambda\!\!\!\!\!\sum_{Q\in\{L,N\}}\int_{0}^{\frac{\pi}{2}}\bigg(1-
\!\!\frac{1}{\left(1\!+\!\frac{s}{N_Q \left(\tan^2{t}+H^2\right)^{\frac{\alpha_Q}{2}} }\right)^{N_Q}}\bigg)\\\notag 
&\quad \quad \quad \quad P_Q\left(\sqrt{\tan^2{t}+H^2}\right)\frac{\sin t}{\cos^3 t}\ dt \bigg).
\end{align}
Finally, by applying Gaussian-Chebyshev approximation the expression in (\ref{Lap}) can be obtained, and the 
proof is complete. 
\end{IEEEproof}
\begin{Remark}
From Lemma $1$ and the corresponding proof, it can be found that the Laplace transform of the interferences 
in the multi-cell pinching-antenna system is the same as that in the conventional fixed antenna system. 
Thus, it can be concluded that the application of pinching-antennas will not cause more severe interferences. 
\end{Remark}
\subsection{Outage probability}
\begin{Lemma}
Given the distance denoted by $d_0$ from the activated pinching-antenna to the typical user in the typical cluster, 
the conditional outage probability can be expressed as follows:
\begin{align}
  P_{out}(d_0)=1-\sum_{B \in \{L,N\}}\sum_{j=0}^{N_B-1}P_B(d_0)\frac{(-\omega)^j}{j!}\bar{\mathcal{L}}^{(j)}(\omega),
\end{align}
where $\epsilon=2^{\bar{R}}-1$, $\omega=N_B\epsilon d_0^{\alpha_B}$, $\bar{\mathcal{L}}(\omega)=\mathcal{L}_I(\omega)e^{-w\xi}$ and 
$\bar{\mathcal{L}}^{(j)}(\omega)$ is the $j$-th derivative of $\bar{\mathcal{L}}(\omega)$. 
\end{Lemma}

\begin{IEEEproof}
By noting that $h_0$ could be a LoS or a NLoS link,  $P_{out}(d_0)$ can be written as 
\begin{align}
   P_{out}(d_0)=\sum_{B \in \{L,N\}}P_B(d_0)\underset{P_{out}(d_0,Q)}
   {\underbrace{\mathbb{P}\left(|g_0|^2 < d_0^{\alpha_B} \epsilon (I+\xi) | d_0, B\right)}}.
\end{align}
Based on the fact that $|g_0|^2$ is Gamma distributed, $P_{out}(d_0,Q)$ can be expressed as follows: 
\begin{align}
 P_{out}(d_0,Q)=1-\sum_{j=0}^{N_B-1}\frac{\omega^j}{j!}\mathbb{E}\left\{(I+\xi)^je^{-\omega(I+\xi)}\right\}
\end{align}
Furthermore, according to the Leibniz integral rule, we have
\begin{align}
 P_{out}(d_0,Q)=1-\sum_{j=0}^{N_B-1}\frac{(-\omega)^j}{j!}\bar{\mathcal{L}}^{(j)}(\omega).
\end{align}
Finally, by summing up the two cases for $B=L$ and $B=N$, the proof can be complete. 
\end{IEEEproof}

\begin{Remark}
Note that the calculation of the derivatives of $\bar{\mathcal{L}}(\omega)$ can be carried out in an iterative way as
follows \cite{sysmmwave2018}: 
\begin{align}
   \bar{\mathcal{L}}^{(j)}(\omega)=\sum_{i}^{j-1}{j-1 \choose i}\zeta^{(j-i)}(\omega)\bar{\mathcal{L}}^{(i)}(\omega),
\end{align}
where $\zeta^{(j)}(\omega)$ can be expressed as follows: 
\begin{align}
  \zeta^{(j)}(\omega)&=-\mathbf{1}(\xi)+\frac{\pi^3\lambda}{2K}\sum_{Q\in\{L,N\}}\sum_{k=1}^{K}\sqrt{1-\theta_k^2}
  \frac{\sin{\phi_k}}{\cos^3{\phi_k}}\\\notag
  &\quad \quad \cdot \frac{P_Q\left(\tan^{2}\phi_k + H^2\right)(-1)^j(N_Q+j-1)!}{(N_Q-1)!N_Q^j(\tan^2 \phi_k +H^2)^{\frac{\alpha_Q j}{2}}}\\\notag
  &\quad \quad \cdot \left(1\!+\!\frac{s}{N_Q \left(\tan^2{\phi_k}+H^2\right)^{\frac{\alpha_Q}{2}} }\right)^{-N_Q-j}.
\end{align}
\end{Remark}

By noting that the typical user is uniformly distributed in the $0$-th cluster and served by its nearest 
pinching-antenna, it is not hard to obtain the expression for the outage probability of the typical user
as highlighted in the following Lemma. 
\begin{Lemma}
The outage probability $P_{out}$ can be expressed as follows:
\begin{align}
P_{out}=\frac{2}{\pi\mathcal{R}^2}\sum_{n=1}^{\bar{N}_P}\int_{A^{L}_n}^{A^{U}_n}\int_{0}^{\sqrt{\mathcal{R}^2-x^2}}
P_{out}(d_{0,n}(x,y))\ dxdy,
\end{align}
where $d_{0,n}(x,y)=\sqrt{(x-\frac{L}{\bar{N}_{P}-1}(n-\frac{\bar{N}_{P}+1}{2})
)^2+y^2+H^2}$, $A^L_n=-\mathcal{R}$ for $n=1$ and $A^L_n=\frac{L}{\bar{N}_P-1}\left(n-\frac{\bar{N}_P}{2}-1\right)$ for 
$2\leq n\leq\bar{N}_P$, $A^U_n=\frac{L}{\bar{N}_P-1}\left(n-\frac{\bar{N}_P}{2}\right)$ for $1\leq n \leq \bar{N}_P-1$ and 
$A^U_n=\mathcal{R}$ for $n=\bar{N}_P$.
\end{Lemma}

\begin{Remark}
It is not hard to find that as the number of preset locations of the pinching antennas, i.e.$\bar{N}_P$, increases, 
the outage probability decreases. Because increasing $\bar{N}_P$ can help the user to find a much closer position for 
activating a pinching-antenna to serve it, which is beneficial for resulting in a LoS link, as well as a lower 
path loss. Therefore, it is interesting to investigate two special cases which provide the upper bound and lower bound of
$P_{out}$, respectively:
\begin{itemize}
  \item Upper bound: $\bar{N}_P=1$, this case degrades to conventional fixed-antenna case where the antenna is fixed at the center of each cluster. 
   In particular, the upper bound can be expressed as follows: 
   \begin{align}
   P_{out}^{\text{upper}}=\frac{2}{\mathcal{R}}\int_{0}^{\mathcal{R}}P_{out}(\sqrt{r^2+H^2})r\ dr.
   \end{align}
  \item Lower bound: $\bar{N}_P \rightarrow \infty$, in this case, the pinching antenna can be activated at any 
   location of waveguide, the lower bound of the outage probability can be expressed as follows:
   \begin{align}
  P_{out}^{\text{lower}}&=\frac{4}{\pi \mathcal{R}^2}\int_{\frac{L}{2}}^{\mathcal{R}}\int_{0}^{\sqrt{\mathcal{R}^2-x^2}}
  P_{out}(d_{0,n}(x,y))\ dxdy\\\notag
  &+\frac{4}{\pi \mathcal{R}^2}\int_{0}^{\frac{L}{2}}\int_{0}^{\sqrt{\mathcal{R}^2-z^2}}P_{out}(\sqrt{y^2+H^2})dydz.
\end{align}
\end{itemize}
\end{Remark}

\begin{Remark}
Based on the results shown in Lemma $3$, the ergodic rate achieved by the typical user
can be evaluated as follows: 
\begin{align}
  R_{\text{Erg}}=\frac{1}{2}\int_0^{\infty}\frac{1-P_{out}}{1+\epsilon}\ d \epsilon.
\end{align}
\end{Remark}

% An interesting question is that how many preset locations do we need to 
% achieve a performance very close to the lower bound? To this end, define 
% the following parameter: 
% \begin{align}
%   \tau = \frac{\log P_{out}}{\log P_{out}^{\text{Lower}}}.
% \end{align}
% It is obvious that $0<\tau<1$, and the larger $\tau$ is, the smaller of gap between $P_{out}$ for a finite
% $\bar{N}_P$ and the lower bound is. Then, the minimum number of preset pinching-antenna locations to guarantee
% the performance gap is not lower than $\bar{\tau}$ can be defined as follows:
% \begin{align}
%  \tilde{N}(\tau)=min\{\bar{N}_P: \tau \geq \bar{\tau}\}.
% \end{align}
\section{Numerical Results}
In this section, numerical results are provided to demonstrate the accuracy of the developed analysis and 
also the superior performance of the pinching-antenna assisted communication system compared to conventional 
fixed-antenna systems.  The parameters are set as follows unless otherwise stated. 
The carrier frequency is $28$ GHz. The transmission bandwidth is 
$100$ MHz. The thermal noise power is $-174$ dBm/Hz. $N_L=3$ and $N_N=2$.

\begin{figure}[!t]
  \centering
      \setlength{\abovecaptionskip}{0em}   % ����ͼƬ������ͼƬ����
      \setlength{\belowcaptionskip}{-2em}   % ����ͼƬ���������ľ���
  \includegraphics[width=2.8in]{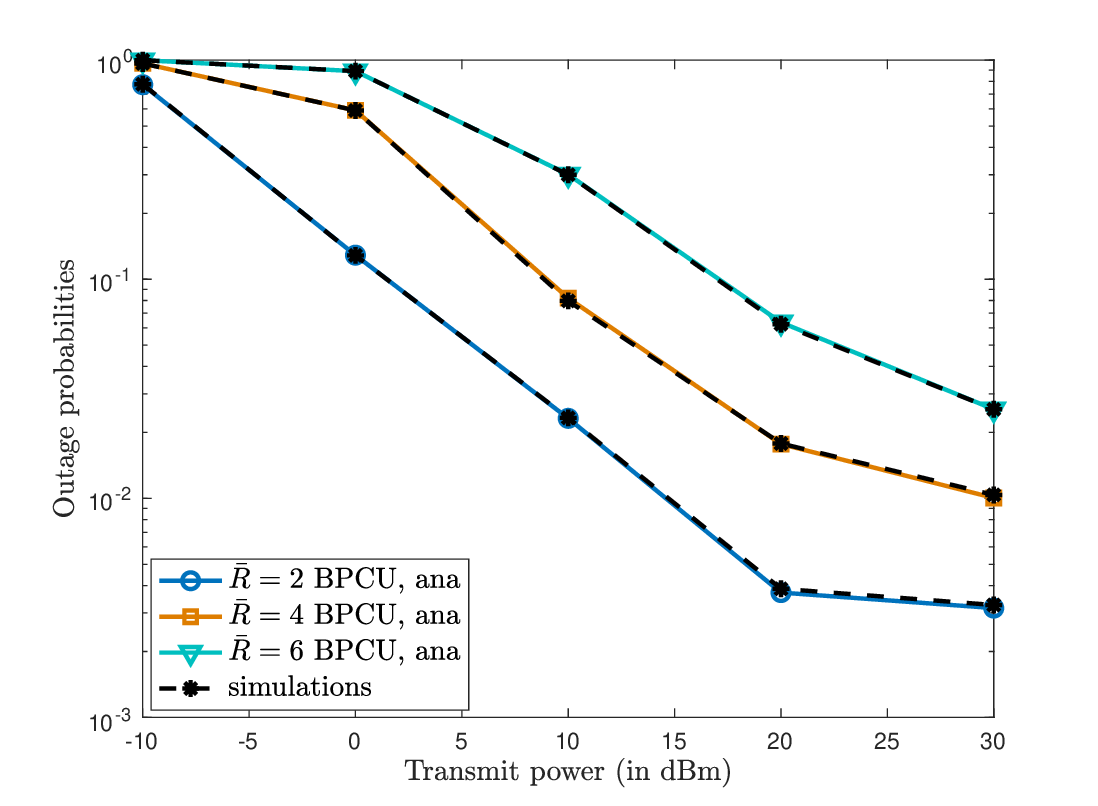}\\
  \caption{Outage probabilities achieved by pinching-antenna system., $\alpha_l=2$, $\alpha_N=3$, $\lambda=10^{-6}$, $\mathcal{R}=20$ m, 
  $L_{w}=10$ m, $\bar{N}_{P}=11$ and $H=3$ m.}\label{accuracy}
\end{figure}

Fig. \ref{accuracy} shows the outage probability achieved by the pinching-antenna system. 
The simulation results are obtained by 
collecting and averaging results from $10^5$ realizations of the point process. 
The analytical results are based on Lemma $3$.
From the figure, it can be observed simulation results perfectly match 
analytical results, which verifies the accuracy of the developed analysis. 

\begin{figure}[!t]
  \centering
      \setlength{\abovecaptionskip}{0em}   % ����ͼƬ������ͼƬ����
      \setlength{\belowcaptionskip}{-2em}   % ����ͼƬ���������ľ���
  \includegraphics[width=2.8in]{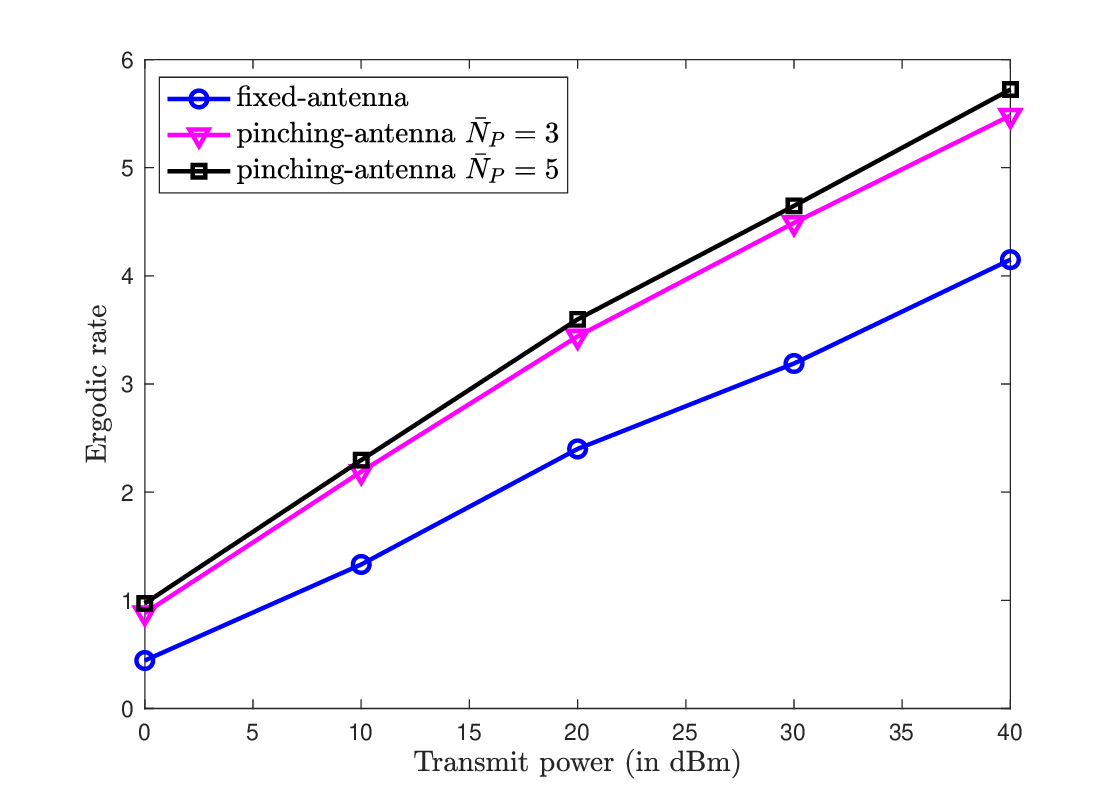}\\
  \caption{Ergodic data rates achieved by pinching-antenna and fixed-antenna systems.$\alpha_l=2$, $\alpha_N=4$, 
  $\lambda=10^{-5}$, $\mathcal{R}=100$ m, 
  $L_{w}=100$ m and $H=4$ m.}\label{ergodic_rate}
\end{figure}

Fig. \ref{ergodic_rate} shows the ergodic data rates achieved by pinching-antenna and conventional fixed-antenna systems. 
From Fig. \ref{ergodic_rate}, it can be observed that the ergodic data rates achieved by pinching-antenna system 
is much higher than those achieved by conventional fixed-antenna system. Another interesting observation is that 
such a performance gain can be obtained with only a few preset positions for activating pinching-antennas. For example, when the transmit power is 
$30$ dBm, the ergodic data rate of pinching antenna with $\bar{N}_P=3$ is about $4.3$ bits per channel use (BPCU), 
while that of fixed-antenna system
is only about $3$ BPCU.

\section{Conclusions}
In this letter, a stochastic geometry based performance analysis on multi-cell pinching-antenna systems with 
spatially distributed waveguides has been provided, which takes the effect of blockage into consideration.
Closed-form expression for the achieved outage probability has been obtained. It has been shown that 
with only a small number of preset pinching-antenna locations where pinching-antennas can be activated, 
pinching-antenna system can achieve much better performance compared to conventional fixed-antenna systems. 

\bibliographystyle{IEEEtran}
\bibliography{IEEEabrv,ref}
\end{document}